\documentstyle[12pt]{article}

\def\laq{\raise 0.4 ex \hbox{$<$}\kern -0.8 em\lower 0.62 ex\hbox{$\sim$}}
\def\gaq{\raise 0.4 ex \hbox{$>$}\kern -0.7 em\lower 0.62 ex\hbox{$\sim$}}
\def\AJ{{\it Ap. J.} }
\def\AJL{{\it Ap. J. Lett.} }

\def\CQG{{\it Class. Quantum Gravity} }
\def\FP{{\it Fortschr. Physik} }
\def\GRG{{\it Gen. Relativity and Gravitation} }

\def\IJMP{{\it Int. J. Mod. Phys.} }

\def\MNRAS{{\it Mon. Not. R. Ast. Soc.} }
\def\NAT{{\it Nature} }
\def\NC{{\it Il Nuovo Cimento} }
\def\NP{{\it Nucl. Phys.} }
\def\PL{{\it Phys. Lett.} }
\def\PR{{\it Phys. Rev.} }
\def\PRL{{\it Phys. Rev. Lett.} }

\def\RMP{{\it Rev. Mod. Phys.} }

\def\vev#1{\langle {#1}\rangle}

\def\frac#1#2{{\textstyle{{#1}\over {#2}}}}

\def\lsim{\mathrel{\rlap{\lower4pt\hbox{\hskip1pt$\sim$}}
    \raise1pt\hbox{$<$}}}
\def\gsim{\mathrel{\rlap{\lower4pt\hbox{\hskip1pt$\sim$}}
    \raise1pt\hbox{$>$}}}
\def\sqr#1#2{{\vcenter{\vbox{\hrule height.#2pt
         \hbox{\vrule width.#2pt height#1pt \kern#1pt
         \vrule width.#2pt}
         \hrule height.#2pt}}}}

\newcommand{\beq}{\begin{equation}}
\newcommand{\eeq}{\end{equation}}
\newcommand{\bea}{\begin{eqnarray}}
\newcommand{\eea}{\end{eqnarray}}

\begin{document}
\titlepage

\begin{flushright}
{DF/IST-1.99\\}
{March 1999\\}
\end{flushright}
\vglue 2cm
	
\begin{center}
{{\bf Compactification, Vacuum Energy and  Quintessence\footnote{
Third Award in the 1999 Essay Competition of the Gravity Research Foundation}\\}
\vglue 1.5cm
{M. C. Bento and O.\ Bertolami\\}
\bigskip
{\it Instituto Superior T\'ecnico,
Departamento de F\'\i sica,\\}
\medskip
{\it Av.\ Rovisco Pais 1, 1096 Lisboa Codex, Portugal\\}
}
\vglue 2cm

\end{center}
\baselineskip=20pt

\centerline{\bf  Abstract}
\vglue 1cm
\noindent
We study the possibility that the vacuum energy density of scalar 
and internal-space gauge fields 
arising from the process of dimensional reduction of higher 
dimensional gravity theories plays the role of quintessence. 
We show that, for the multidimensional 
Einstein-Yang-Mills system compactified on a ${\bf R} \times S^3 \times S^d$ 
topology, there are classically stable solutions such that the observed 
accelerated expansion of the Universe at present 
can be accounted for without upsetting structure formation scenarios 
or violating observational bounds on the vacuum energy density.

\vfill
\newpage

\setcounter{equation}{0}
\setcounter{page}{2}
%%%%%%%%%%%%%%%%%%%%%%%%%%%%%%%%%%%%

\baselineskip=20pt

\section{Introduction}

\indent

\vspace*{0.3cm}
Recently,   strong evidence is emerging that the Universe is 
dominated by a smooth component with an effective negative pressure 
and expanding in an accelerated fashion. 
These findings arise from the study of more than 50 
recently discovered Type IA Supernovae with red-shifts greater than 
$z \ge 0.35$ \cite{Per}. Such studies, carried out by two
different groups \cite{Per, Riess}, lead to the striking result that 
the deceleration parameter 

\begin{equation}
q_{0} \equiv - { \ddot{a} ~a \over \dot{a}^2}~,
\label{1.1} 
\end{equation}
where $a(t)$ is the scale factor, is negative 

\beq
-1~\laq~q_0~<~0~.
\label{1.01}
\eeq
It follows from the Friedmann and Raychaudhuri equations for an 
homogeneous and isotropic geometry that, if the sources driving the 
expansion are  vacuum energy and matter, with equation of state $
p =  \omega \rho,~ -1 \le \omega \le 0$, then 
the deceleration parameter is given by:

\begin{equation}
q_{0} = {1\over 2 } (3 \omega  + 1) \Omega_{M} - \Omega_{\Lambda}~~,
\label{1.3} 
\end{equation}
where $\Omega_{M(V)}$ denotes the energy density of matter (vacuum) 
in units of the critical density. For a Universe where the 
matter component is 
dominated by non-relativistic matter or dust, $\omega = 0$,
and therefore the combination $\Omega_{M} \sim 0.4$ and 
$\Omega_{\Lambda} \sim 0.7$ seems observationally favoured.
Of course, the value $\Omega_{\Lambda} \sim 0.7$, although
consistent with observation (see Ref. \cite{Ber1} for a list of the important
constraints), implies  a quite unnatural
fine tuning of parameters if it arises from any known particle physics setting
(see Refs. \cite{Weinberg} for a thorough review and Refs. 
\cite{Ber1, Ber2}  for possible  connections with fundamental 
symmetries like
Lorentz invariance and S-duality in string theories). 
Furthermore, $\Omega_{M}$ and $\Omega_{\Lambda}$ of the same order suggests
that we live in a rather special cosmological period.

While the most straightforward candidate for a smooth component is a 
cosmological constant, a plausible alternative is a dynamical vacuum 
energy, or ``quintessence''. Suggestions along these lines have been proposed 
a long time ago \cite{Ber3}, although yielding a vanishing  deceleration 
parameter. A number of quintessence models  have 
been put forward, the most popular of which invoke a scalar field with 
a very shallow potential, which until recently was overdamped in its 
evolution by the expansion of the Universe, allowing for its energy density 
to be smaller than the radiation energy density at early times, such that 
at present $\Omega_{M}~\laq~\Omega_{\Lambda}$ \cite{Zlat, Stein}.
It was also shown that scalar fields
with an exponential type potential can, under conditions, 
render a negative $q_0$ \cite{Peeb, Ferreira}. Other suggestions 
include the string theory dilaton together with 
gaugino condensation \cite{Binetruy}, an axion with an almost massless
quark \cite{Kim}, a time-dependent vacuum energy induced by $\it D$-particle
recoil \cite{Ellis}, etc. However interesting, most of 
these suggestions necessarily involve a quite severe fine tuning of parameters
\cite{Kol}.
This fact calls for constructions that allow for a negative deceleration 
parameter using sources of quintessence that ideally do not 
require a potential. In this context, it has been shown
that a  scalar field coupled with gravity non-minimally, namely a 
self-interacting Brans-Dicke type field with a negative coupling, 
can be used for this purpose \cite{BerMar}.

In this work, we study the possibility that scalar fields
arising from the process of 
dimensional reduction of higher dimensional gravity theories, together 
with internal gauge fields, play the role of quintessence. The stability 
of the required compactification of the extra dimensions is
related with the dynamics of these fields. Classical and quantum 
stabilility  depends on the existence of minima of the relevant potential 
that are classically or, at least, semiclassically stable.
We show that, for the multidimensional 
Einstein-Yang-Mills system \cite{KRT,BKM,BFM}, the  cosmological framework 
following from demanding that compactifying solutions 
are classically stable can also be 
used to drive an accelerated expansion at present, 
at the expense of the contribution of higher dimensional fields. In fact, the
stability of compactification requires fine tuning the higher-dimensional
cosmological constant meaning that, in this respect, our proposal is
afflicted with the same difficulty of other quintessence models.
Nevertheless, the most advantageous aspects of our setting are that it 
rests on the fruitful ground 
of the Einstein-Yang-Mills system and, therefore,  there is no need to
postulate {\it ad hoc} potentials and also that the dimensional reduction
procedure determines, via the theory of symmetric fields, the cosmological 
model unambiguously. 
Thus, in a single framework, the issues of compactification 
and accelerated expansion of the Universe are related and  
the ground-state energy of fields emerging 
from the compactification scenario can actually be regarded as a 
consistent quintessence candidate, a scenario that we choose to
call ``quintessential compactification''.

\section{The Generalized Kaluza-Klein Model}

\vspace*{0.3cm}

Compactification is a crucial step in rendering multidimensional 	
theories of unification, such as generalized Kaluza--Klein theories, 
Supergravity and Superstring theories, consistent with our four-dimensional 
world. Phenomenology requires that the extra
dimensions  are stable and Planck size (see, however, 
\cite{Arkani} for a different proposal  concerning this issue). 
A necessary condition
for the stability of the extra dimensions is the presence of matter 
with repulsive stresses to counterbalance gravity. Magnetic
monopoles \cite{DSS}, Casimir forces \cite{AC} and Yang-Mills fields
\cite{KRT,BKM} have been suggested for this purpose. The case of 
Yang-Mills fields is
particularly interesting as it illustrates the  importance of considering
non-vanishing internal as well as external-space components 
of the gauge fields. 
Indeed, as shown  in Ref. \cite{BKM}, it is this feature that renders 
compactifying solutions classically as well as semiclassically stable. 
Moreover, it was shown in Ref. \cite{BFM}, using the quantum cosmology 
formalism, that, for the Einstein-Yang-Mills system, compactifying 
solutions with non-vanishing 
external-space components of the gauge field are correlated with the 
expansion of the Universe. This implies that, for expanding universes, it is
more likely that stable compactification solutions arise.

Following \cite{KRT,BKM}, we consider an $SO(N)$ gauge field with 
$N \ge 3 + d$ in $D = 4 + d$ dimensions and
an homogeneous and (partially) isotropic spacetime in a ${\bf R} 
\times S^3 \times S^d$ topology. The relevant
coset compactification splitting of the $D$-dimensional spacetime $M^D$
is the following      

\begin{equation}
M^D = {\bf R} \times G^{{\rm ext}}/H^{{\rm ext}} \times G^{{\rm int}}/
H^{{\rm int}}~,
\label{2.1}
\end{equation}
where $\bf R$ denotes the timelike direction, $G^{\rm ext(int)} = SO(4) 
(SO(d+1))$ and $H^{{\rm ext(int)}} = SO(3) (SO(d))$ are
respectively the homogeneity and isotropy groups in 3($d$) dimensions.
For the multidimensional 
Einstein-Yang-Mills model we consider, the gauge field has only  
time-dependent spatial components on the 3-dimensional 
physical space.

The model is derived from the generalized Kaluza-Klein action: 
\begin{equation}
S[\hat g_{\hat\mu\hat\nu}, \hat A_{\hat\mu}, \hat \chi] = 
S_{{\rm gr}}[\hat g_{\hat\mu\hat\nu}]
 + 
S_{{\rm gf}}[\hat g_{\hat\mu\hat\nu}, \hat A_{\hat\mu}]
 + 
S_{{\rm inf}}[\hat g_{\hat\mu\hat\nu},   \hat \chi]~
\label{2.2}
\end{equation}
with 
\begin{eqnarray}
S_{{\rm gr}}[\hat g_{\hat\mu\hat\nu}] & = & 
{1 \over 16 \pi \hat k} \int_{M^D} d \hat{x} 
\sqrt{-\hat g}~(\hat R - 2 \hat \Lambda) ~, \label{2.2a}
\\
S_{{\rm gf}}[\hat g_{\hat\mu\hat\nu}, \hat A_{\hat\mu}]
& = & 
{1 \over 8\hat{e}^2} 
\int_{M^D} d \hat{x} 
\sqrt{-\hat g}~{\rm Tr} \hat F_{\hat\mu\hat\nu} 
\hat F^{\hat\mu\hat\nu} ~, \label{2.3a} \\
S_{{\rm inf}}[\hat g_{\hat\mu\hat\nu},   \hat \chi]
& = & 
- \int_{M^D} d \hat{x} 
\sqrt{-\hat g}~\left[{1 \over 2} 
\left(\partial_{\hat\mu} \hat \chi\right)^2 
+ \hat U \left(\hat \chi\right)
\right] ~, 
\label{2.3} 
\end{eqnarray}
where $\hat g$ is $\det 
\left(\hat g_{\hat\mu\hat\nu}\right)$, 
$\hat g_{\hat\mu\hat\nu}$ is the $D$-dimensional metric, 
$\hat R$, $\hat e$, $\hat k$ and $\hat \Lambda$ are, 
respectively, the scalar curvature, gauge coupling, 
 gravitational and cosmological constants 
in $D$ dimensions. In addition, the following 
field variables are defined in $M^D$: 
$\hat F_{\hat\mu\hat\nu} = 
\partial_{\hat\mu} \hat A_{\hat\nu} - 
\partial_{\hat\nu} \hat A_{\hat\mu} + 
\left[ \hat A_{\hat\mu}, \hat A_{\hat\nu}\right]$ 
is the gauge field strength, 
$\hat A_\mu$ denotes the components of the 
gauge field and  $\hat \chi$ is the inflaton, 
responsible for the inflationary expansion of the external 
space. The  inflaton potential,  $\hat U(\hat \chi)$, is taken to be 
bounded from below,  having a global minimum so that $\hat U_{{\rm min}} = 0$.

We consider vacuum solutions where the splitting of internal and
external dimensions of spacetime  corresponds 
to a factorization in a product of spaces

\begin{equation}
M^D =  M^4 \times K^d~~,
\label{2.4}
\end{equation}
$M^4$ being the four-dimensional Minkowski  spacetime, 
$K^d$ a Planck-size $d-$dimensional compact space. We  assume that  
$M^{D} = {\bf R} \times S^3 \times S^d$,
where $S^3$ and $S^d$ are $3$ and $d$-dimensional spheres.

The  spatially homogeneous and (partially) 
isotropic field configurations relevant for our cosmological model are 
symmetric under the action of the group 
$G^{{\rm ext }} \times  
G^{{\rm int}}$. The 
following realization of $M^D$ can then be constructed \cite{BKM}

\begin{eqnarray}
M^{D} & = & {\bf R} \times SO(4)/SO(3) \times SO(d+1)/SO(d) \nonumber \\
& =& {\bf R} \times [SO(4)\times SO(d+1)]/[SO(3) \times SO(d)] ~.
\label{2.7}
\end{eqnarray}

The metric corresponding to the D-dimensional spacetime is given by

\beq
ds^2=-{\tilde N}^2(t) dt^2 + {\tilde a}^2 d\Omega_3^2 + b^2(t) d\Omega_d^2~,
\label{2.11b}
\eeq
where $\tilde a(t)$ and $b(t)$ are the scale factors of $S^3$ and $S^d$ 
respectively, and $\tilde N(t)$ is the lapse function.

The remaining field configurations associated with the above geometry,  
described in Ref. \cite{BKM}, are built using the theory of symmetric fields
(see eg. \cite{bmpv} and references therein). 
Substituting the corresponding Ans\"atze  into the action 
(\ref{2.2}) and performing the conformal transformations 
\begin{eqnarray}
\tilde N^2(t) & = &  \left[{\vev{b} \over b(t)}\right]^d N^2 (t)~, 
\label{2.12a} \\ 
\tilde a^2 (t) & = & 
\left[{\vev{b} \over b(t)}\right]^d a^2 (t)~, 
\label{2.12b}
\end{eqnarray}
where $\vev{b}$ is the vacuum expectation value   of $b(t)$, we obtain a 
one-dimensional effective reduced action \cite{BKM}:

\begin{eqnarray}
S_{\rm eff} & =& 
16 \pi^2 {\large \int} dt N a^3 \left\{ -{3 \over 8\pi k} 
{1 \over a^{2}} \left[{\dot{a} \over N}\right]^2 + 
{3 \over 32\pi k} {1 \over a^{2}} + {1 \over 2} 
\left[
{\dot{\psi} \over N}\right]^2  + 
{1 \over 2}\left[{\dot{\chi} \over N}\right]^2 \right. \nonumber \\
&+&  \left.  e^{d\beta\psi} {3 \over 4e^2} {1 \over a^{2}} 
\left({1 \over 2}\left[{\dot{f_0} \over N}\right]^2 
+ {1 \over 2}\left[{ {\cal D}_t {\bf f} \over N}\right]^2\right) 
+ e^{-2\beta\psi} {d \over 4e^2} {1 \over \vev{b}^{2}} 
{1 \over 2}\left[{{\cal D}_t {\bf g} \over N}\right]^2 - 
W\right\},
\label{2.13}
\end{eqnarray}
where $k = \hat k /v_d \vev{b}^d$ is Newton's constant, 
$e^2 = \hat e^2/v_d \vev{b}^d$, $e$ being the gauge coupling, 
$\beta = \sqrt{16 \pi k / d(d+2)}$ and $v_d$ is the the volume 
of $S^d$ for $b=1$. Moreover, we have set
$\psi \equiv  \beta^{-1} \ln (b/\vev{b})$ and 
$\chi \equiv  \sqrt{v_d \vev{b}^d} \hat \chi$. The dots denote 
time derivatives and ${\cal D}_t$ is the covariant derivative with 
respect to the remaining  $SO(N-3-d)$ gauge field $\hat B(t)$ in {\bf R}:
\begin{equation}
{\cal D}_t {\bf f}(t) = {d \over d t} {\bf f(t)} + 
\hat B(t) {\bf f}(t)~~, ~
{\cal D}_t {\bf g}(t) = {d \over d t} {\bf g(t)} + 
\hat B(t) {\bf g}(t)~~.
\label{2.14}
\end{equation}

It is important to point out that $f_0(t), {\bf f} = \left\{ f_p \right\}$ 
represent the gauge 
field components in  4-dimensional physical space-time, while 
${\bf g} = \left\{ g_q \right\}$  denotes the components in  $K^d$, $\hat B$ is an
$(N-3-d)\times (N-3-d)$ antisymmetric matrix 
and $\psi$ is the scalar field emerging from 
the compactification procedure. 

The potential $W$, in (\ref{2.13}), is given by 
\begin{eqnarray}
W & = & e^{-d\beta\psi} \left[ -e^{-2\beta\psi}
{1 \over 16\pi k} {d(d-1) \over 4} {1 \over \vev{b}^{2}} + 
e^{-4\beta\psi} {1 \over \vev{b}^{4}} {d(d-1) \over 8 e^2} 
V_2 ({\bf g})
+  {\Lambda \over 8\pi k} + U(\chi) 
\right] \nonumber \\ & +  & 
e^{-2\beta\psi} {1 \over (a\vev{b})^{2}} {3d \over 32 e^2} 
({\bf f}\cdot {\bf g})^2 + e^{d\beta\psi} {3 \over 4 e^2 a^4} 
V_1(f_0, {\bf f})~~,
\label{2.15}
\end{eqnarray}
where 
$\Lambda = v_d \vev{b}^d \hat \Lambda$, $U(\chi) = v_d \vev{b}^d 
\hat U \left(\hat \chi / \sqrt{v_d b^d_0}\right)$ and 
\begin{eqnarray}
V_1 (f_0, {\bf f}) & =  & {1 \over 8} 
\left[ \left( f_0^2 + {\bf f}^2 - 1 \right)^2 + 
4 f_0^2 {\bf f}^2 \right]~~,
\label{2.16a} \\
V_2 ({\bf g}) & =  & {1 \over 8} \left( {\bf g}^2  - 1\right)^2~~,
\label{2.16b}
\end{eqnarray}
are related with the external and internal components of the gauge fields, 
respectively.
Variables $N$ and $\hat B$ are Lagrange multipliers associated 
with the symmetries of the effective action (\ref{2.13}). 
The lapse function $N$ is related to 
the invariance of $S_{{\rm eff}}$ under arbitrary time 
reparametrizations, while $\hat B$ is connected with the 
local remnant $SO(N-d-3)$ gauge invariance. Without loss of generality 
the gauge $N = 1$ will be used in what follows. The equations of motion 
for the physical variables $a, \psi, \chi, f_0, {\bf f}, {\bf g}$ 
can be found in Ref. \cite{BKM}.

The Friedmann equation and the 
equation of motion for the field  $\psi$, relevant for our quintessence
proposal are the following:

\begin{equation}
\left({\dot{a} \over a}\right)^2 = - {1 \over 4 a^2} 
+ {8 \pi k \over 3} \left[{\dot{\psi}^2 \over 2} + W(a, \psi)+\rho\right]~~,
\label{2.17}
\end{equation}

\begin{equation}
\ddot{\psi} + 3 \left({\dot{a} \over a}\right) \dot{\psi}
+ {\partial W \over \partial \psi} = 0~~.
\label{2.18}
\end{equation}
Notice that  we have added a term corresponding to the matter energy density 
since its contribution is quite important for the late time Universe, 
$\rho=\rho_0 (\frac{a_0}{a})^3$, 
where $\rho_0$ and $a_0$ are the matter energy density and the scale factor at
present, respectively.
The compactification scenario we envisage involves static vacuum 
configurations of the gauge and inflaton fields:
\begin{equation}
f_0 = f_0^v, ~{\bf f} = {\bf f}^v, ~{\bf g} = {\bf g}^v = {\bf 0}, ~
\chi = \chi^v~~.
\label{2.18a}
\end{equation}
We also assume that ${\bf f}$ and ${\bf g}$ are 
orthogonal and that $U(\chi^v) = 0$. For simplicity, we use the notation 
$v_1 \equiv V_1 (f_0^v, {\bf f}^v)$ and 
$v_2 \equiv V_2 ({\bf g}^v)= {1 \over 8}$ in what follows. The potential 
(\ref{2.15}) simplifies then to
\beq
W  =  e^{-d\beta\psi} \left[ -e^{-2\beta\psi}
{1 \over 16\pi k} {d(d-1) \over 4} {1 \over \vev{b}^{2}} + 
e^{-4\beta\psi} {1 \over \vev{b}^{4}} {d(d-1) \over 8 e^2} 
v_2 +  {\Lambda \over 8\pi k}\right] 
+  e^{d\beta\psi} {3 \over 4 e^2 a^4} v_1~~,
\label{2.19}
\eeq
where the last term arises from the contribution of the external-space 
components of the gauge field and clearly represents the contribution
of radiation for the energy density of the Universe.

As discussed in Ref. \cite{BKM}, different 
values for the cosmological constant $\Lambda$ correspond to different 
compactification scenarios. Indeed, for $\Lambda>c_2/16\pi k$, where   
$c_2=[(d+2)^2(d-1)/(d+4)]e^2/16v_2$, 
there are no compactifying solutions and for

\beq
{c_1 \over 16\pi k}<\Lambda< {c_2 \over 16\pi k}~~,
\label{2.20}
\eeq
with $c_1=d(d-1)e^2/16v_2$, a compactifying solution exists which is 
classically
stable but semiclassically unstable. On the other hand, a value of 
$\Lambda<c_1/16\pi k$ implies that the effective 
4-dimensional cosmological constant, 
$\Lambda^{(4)}= 8\pi k W(a\rightarrow\infty,\psi)$, is negative.
As $\Lambda^{(4)}$ must satisfy the order of magnitude observational bound
\beq
\Lambda^{(4)} \approx 10^{-120} {1 \over 16\pi k}~~,
\label{2.21}
\eeq
we consider the following fine-tuning of the multidimensional cosmological 
constant $\Lambda=c_1(1+\delta)/16\pi k$, where $\delta$ is clearly 
proportional to $\Lambda^{(4)}$. 
On the other hand, since we 
are interested in compactifying solutions, for which $\psi\approx 0$, 
we choose $\Lambda$ such that $\psi=0$ corresponds to the absolute minimum of 
(\ref{2.19}), where  $\vev{b}^2 = {16\pi k v_2 /e^2}$. Hence 
\beq
\Lambda={d(d-1) \over 16\vev{b}^2}~(1 + \delta)~~.
\label{2.22}
\eeq
Substituting (\ref{2.22}) in (\ref{2.19}), we obtain, 
for large $a$ (implying that  the radiation term can be neglected)
\beq
W={d(d-1)\over 128 \pi k \vev{b}^2}\delta~~.
\label{2.23}
\eeq

\vspace{0.3cm}
Of course, a non-vanishing $\delta$ introduces a semiclassical instability
in our compactification solution; however, the decay rate of the compactified 
vacuum is such that the decompactification time is much greater, by many orders
of magnitude, than the age of the Universe.

Naturally, compactification occurs prior inflation, which is
driven by the potential $U(\chi)$ of the inflaton, $\chi$. It should  
be pointed out that the inflaton itself could, via a suitable
choice of its potential or by relaxing the condition $U(\chi^v) = 0$, 
be at the origin of a late accelerated expansion. 
This possibility has been proposed some time ago and was 
called ``deflation''
\cite{Spokoiny} or ``quintessential inflation'' in a more recent version 
of the idea \cite{PeebVil}. This is certainly an interesting suggestion  
that can actually be implemented in the context of our model. 
Interestingly, quintessential inflation can be, in principle, detected. 
Indeed, as discussed in Ref. \cite{PeebVil}, a distinct feature of 
quintessential inflation is the form of its gravitational wave spectrum, 
which although inacessible to gravity wave detectors under construction, 
such as LIGO and VIRGO, may be within reach for a future generation of 
detectors.  
\vspace{1cm}

\section{The Deceleration Parameter in Quintessential Compactification}

\vspace*{0.3cm}

Let us now compute the deceleration parameter for late times in the context 
of our quitessential compactification scenario. 
We differentiate the
Friedmann equation (\ref{2.17}) and substitute the resulting term in 
$\ddot\psi$ by eq. (\ref{2.18}), to obtain

\beq
\ddot a = 
{8 \pi k \over 3} a \left[{\dot{\psi}^2 \over 2} + W(a, \psi)\right] - 
{4\pi k\over 3 a^2}\rho_0 a_0^3 + 
{4 \pi k \over 3 \dot a } a^2 \left[{-3 H \dot{\psi}^2 } + 
{\partial W\over \partial a} \dot a\right]~~,
\label{2.24}
\eeq
where $H={\dot a \over a}$.
Neglecting the last term, a very good approximation since 
${\partial W\over \partial a}\sim a^{-5}$, we have

\beq
\ddot a = - {8 \pi k \over 3} a \left[{\dot{\psi}^2 } - W(a, \psi) + 
{\rho_0 a_0^3\over 2 a^3}\right]~~.
\label{2.25}
\eeq

For $\psi = 0~$\footnote{Assuming that $\psi$ oscillates 
around the minimum, i.e. $W\sim \psi^2$, we then have, according to 
the virial theorem, $<\dot{\psi}^2> = <W>$. In this case, 
we would obtain a positive $q_0$. However, since 
the minimum is very deep and the time scale of the problem is the 
Planck time, it is reasonable to assume that $\psi$ has had enough 
time to settle at the minimum of the potential.}, we get for the 
deceleration parameter, substituting (\ref{2.23}) in (\ref{2.25}) 
and setting  $\dot\psi=0$

\beq
q={-{a^2\over \vev{b}^2}\hat \delta + 
{4\pi k \over 3}{\rho_0 a_0^3\over a }\over -\frac{1}{4} + 
{a^2\over \vev{b}^2 }\hat\delta + {8\pi k\over 3}{\rho_0 a_0^3\over a}}~~,
\label{2.26}
\eeq
where $\hat \delta={d(d-1)\over 48}\delta$. Given the bound (\ref{2.21}), 
then $\delta=\delta_0 10^{-120}$, where $\delta_0$ will be computed below. 
On the other hand, taking  the gauge coupling $e \sim 0.3$, 
the radius of the compact manifold 
$K^d$ is about $10$ times greater than the Planck size \footnote{It is 
interesting to notice that, if $\vev{b}$ were much greater than the 
Planck size, as suggested in \cite{Arkani}, it  would render our quitessence 
proposal untenable as the order of magnitude of $\delta$ is 
determined by (\ref{2.21}).}, and 
$\left({a_0 \over \vev{b}}\right)^2 = \alpha_{0} 10^{120}$, where 
$\alpha_{0}$ is an order one constant.
Hence, we obtain for the deceleration parameter at present

\beq
q_0={-\delta_1 + {\epsilon\over 2}\over -{1\over 4 }+ \delta_1 + \epsilon}~~, 
\label{2.27}
\eeq
where $\delta_1 \equiv d(d - 1) \alpha_{0} \delta_0 / 48$ and
\beq
\epsilon  \equiv  {8\pi k\over 3}{\rho_0 a_0^2} = 
{320 \pi\over 3} \alpha_0\Omega_M h_0^2~~, 
\label{2.28}
\eeq
where $h_{0}$
parametrizes the observational uncertainty in the Hubble constant, 
$H_0 = 100~h_0~km~s^{-1}~Mpc^{-1}$ and $0.4~\laq~h_{0}~\laq~0.7$.

In order to obtain a bound on $\delta_0$, we compute the time $t_q$ 
when quintessence states started dominating the dynamics of the Universe. 
The value of $a_q=a(t_q)$ can be
estimated equating the contributions of $W$ and $\rho(a_q)$. From 
the observational bound $\Omega_M~\laq~0.3$ \cite{nbahcall}, we obtain:

\beq
\alpha^3~\delta_0~\laq~{1536~\pi \over d(d - 1)}~h_{0}^2~~,
\label{2.29}
\eeq 
where $\alpha\equiv {a_q\over a_0}$.  
Since the red-shift of the supernova data used to infer the accelerated 
expansion of the Universe is $z \ge 0.35$, then  
$\alpha~\le~0.74$ and, for $d = 7$ and $h_{0} = 0.5$, we get

\beq
\delta_1~\laq~71~,
\label{2.30}
\eeq
which implies, for e.g.  $\delta_1 = 70$, that

\beq
q_0 = -~0.56~,
\label{2.31}
\eeq
which sits inside the most likely region of values for $q_0$, as revealed by
observational data \cite{Per, Riess}.
 
Notice that, with  $\alpha~\le~0.74$, the quintessence 
domination period is rather recent in the history of the Universe and, hence, 
structure formation scenarios, for which the condensation period is 
$z_{c}~\gaq~10$, given the most recent Hubble deep field surveys, 
are unaffected by our quintessence proposal 
provided the vacuum energy density is not too 
large. It is easy to show that this is indeed the case as, even for $z\approx 0.35$, the time when 
quintessence starts dominating the dynamics, 
$\Omega_{V}(z \ge 0.35) \simeq 0.5$, consistent with bounds arising 
from anthropic considerations which  imply that, for a flat Universe ( which is not 
the case of our model),  a vacuum energy no greater  than about 
$\pi (1 + z_{c})^3 \rho_{0}$ does not prevent 
gravitational condensation \cite{Weinberg}. At present, one should expect 
$\rho_V~\laq~3~\rho_{0}$ \cite{Martel}. Our model is compatible with 
this bound and also with the upper limit arising from 
gravitational lensing studies, namely that, at present, $\Omega_{V} < 0.75$
\cite{Fukugita, Maoz}.

A distinct feature of the model is that, in spite of having a 
closed topology,  a phase of accelerated 
expansion can take place. Indeed, writing the Friedmann equation as
\beq
{\dot a}^2 + V(a) = - {1\over 4}~~,
\label{2.32}
\eeq
with

\beq
V(a) =- {a^2\over\vev{b}^2 }\hat\delta  - {8 \pi k\over 3} 
{\rho_0 a_0^3\over a}~~,
\label{2.33}
\eeq
we see that accelerated expansion can be achieved provided

\beq
V(a_M)~ < ~-{1\over 4}~~,
\label{2.34}
\eeq
where $a_M$ is the maximum of $V(a)$. This requirement implies 
the condition

\beq
4~\delta_1~\mu^2 + {\epsilon \over \mu} - 1 > 0~~,
\label{2.35}
\eeq
where $\mu \equiv {a_M \over a_0}$. It can easily be verified that, for 
$\delta_1 = 70$ and $\alpha_{0} = {\cal O}(1)$, this inequality holds for 
any $\mu$, hence representing a consistency check for our proposal. 
Clearly, our model presents, however brief, a coasting period where 
$a \sim constant$.

\section{Discussion and Conclusions}

\vspace*{0.3cm}

In this work, we have proposed a cosmological 
model based on the multidimensional Einstein-Yang-Mills system, 
compactified on ${\bf R} \times S^3 \times S^d$ spacetime. We have shown that 
the very fine tuning on the higher dimensional cosmological
constant needed  for stable compactifying solutions and   
rendering  the vacuum energy density consistent with observational 
bounds, can also 
account for the observed accelerated expansion of the Universe, 
for reasonable values of the model parameters. Specifically, we 
obtain a deceleration parameter  $q_0 = -~0.56$. This is achieved 
via the vacuum contribution of a scalar field, very much like in 
the so-called quintessence 
scenarios but, in our model, an internal gauge field is also 
involved and these fields arise from the compactification process via the 
dimensional reduction procedure. 
Furthermore, we have shown that, since the quintessence 
domination period is quite recent in the history of the Universe, 
known scenarios for structure formation remain unaffected by 
our quintessence proposal and bounds on the vacuum energy density are 
respected. 

Finally, it is interesting to point out that our setting
allows, quite naturally, for  a quintessential inflationary extension, and although
we have chosen here to study the ``minimal'' version of the
cosmological model arising from the multidimensional Einstein-Yang-Mills
theory, further work on ``quintessential compactification - inflation'' 
follows  immediately from the model.

\end{document}